\documentclass[final,3p,times]{elsarticle}

\usepackage{CJK}

\usepackage{graphicx} 
\usepackage{amssymb}
\usepackage{amsmath}
\usepackage{subfigure}
\usepackage{graphicx}
\usepackage{epstopdf}
\usepackage{bibentry}
\usepackage{color}
\usepackage{breqn}
\usepackage{lineno,hyperref}

\allowdisplaybreaks[4]

\journal{}

\begin{document}

\begin{frontmatter}



\title{Semi-rational solutions for the $(2+1)$-dimensional nonlocal Fokas system}



\author[1]{Yulei Cao}
\author[2]{Jiguang Rao}
\author[3]{Dumitru Mihalache}
\author[1]{\corref{cor1} Jingsong He}
\address[1]{Department of Mathematics, Ningbo University,
Ningbo, Zhejiang, 315211, P.\ R.\ China}
\address[2]{School of Mathematical Sciences, USTC, Hefei, Anhui 230026, P.\ R.\ China}
\address[3]{Horia Hulubei National Institute for Physics and Nuclear Engineering,
P.O.B. MG-6, Magurele, RO 077125, Romania}
\cortext[cor1]{e-mail: hejingsong@nbu.edu.cn}

\begin{abstract}
The $(2+1)$-dimensional [$(2+1)d$] Fokas system is a natural and simple extension of the nonlinear Schr\"{o}dinger equation (see  eq. (2) in A. S. Fokas, Inverse Probl. 10 (1994) L19-L22). In this letter, we introduce its $\mathcal{PT}$-symmetric version, which is  called
the $(2+1)d$ nonlocal Fokas system. The $N$-soliton solutions for this system are obtained by using the Hirota
bilinear method whereas
the semi-rational solutions are generated by taking the long-wave limit of a part of
exponential functions in the general expression of the $N$-soliton solution.
Three kinds of semi-rational solutions, namely (1) a hybrid of rogue waves and periodic line waves, (2) a hybrid of lump and breather solutions, and (3) a hybrid of lump, breather, and periodic line waves are put forward and their rather complicated dynamics is revealed.
\end{abstract}


\end{frontmatter}


\section{Introduction}
The study of nonlinear evolution equations (NLEEs) has made fantastic progress, and it is frequently used to model various nonlinear phenomena in physics, chemistry, biology, and even social sciences. The analytical solutions of NLEEs are  necessary for a better understanding of those nonlinear phenomena. Many researchers have made great efforts to find analytical solutions of those equations and a series of powerful techniques have been proposed, such as the Lie group analysis \cite{ml1,ml2}, the inverse scattering transformation (IST) \cite{mj1,mj2}, the Darboux transformation (DT) \cite{hj1,hj2}, the Hirota bilinear method  \cite{yo1,yo2,yo3}, and other techniques \cite{as1,as2,as3,as4,as5,as6}. To date, most of the relevant studies were focused on solitons, breathers, rogue waves (RWs), and lump solutions. However, during the past few years, studies of the dynamics of semi-rational solutions of NLEEs have become an important research field  \cite{jm1,jm2,jm3,jm4}. Generally speaking, comparing to $(1+1)$-dimensional [$(1+1)d$] systems, the methods to solve $(2+1)$-dimensional [$(2+1)d$] systems are much more complicated. Therefore, the study of $(2+1)d$ semi-rational solutions of NLEEs would be much more challenging and meaningful.

Parity-time [$\mathcal{PT}$] symmetry of physical systems has been extensively studied both theoretically and experimentally since the pioneering works of Bender {\it et al.} \cite{BBS1,BBS2,BBS3}, who showed that a wide class of non-Hermitian Hamiltonians having the $\mathcal{PT}$ symmetry can possess entirely real spectra as long as this symmetry is not spontaneously broken. The $\mathcal{PT}$ symmetry is maintained in optics by means of a special balance between gain and loss in the corresponding regions of the optical system \cite{kg1,kg2}. The studies of $\mathcal{PT}$ systems have led to new developments in diverse areas of theoretical and applied physics, including quantum field theories \cite{cm1}, Lie algebras \cite{cm2}, complex crystals \cite{cm3,cm4,cm5}, and optics and photonics \cite{cm6}.

Inspired by the above results of $\mathcal{PT}$-symmetric physical systems, we introduce a new $\mathcal{PT}$-symmetric  equation,
\begin{equation}\label{N1}
\begin{aligned}
iU(x,y,t)_{t}+U(x,y,t)_{xx}+U(x,y,t) \int^{y}_{-\propto}[U(x,y,t)U(-x,-y,t)^*]_x dy^{'} =0,\qquad  \\
\end{aligned}
\end{equation}
which is an extension of the $(2+1)d$ Fokas system \cite{FO}, and thus is called  the $(2+1)d$ nonlocal Fokas system. There are some interesting results on the  $(2+1)d$ Fokas system \cite{we1,we2,we3,we4,we5}, including  solitons, lumps, and line-rogue waves. Instead, we shall focus on  the above newly established nonlocal system, and  construct its $N$-soliton and
semi-rational solutions.

Using the transformations
\begin{equation} \label{N2}
\begin{aligned}
V_{y}=[U(x,y,t)U^{*}(-x,-y,t)]_{x},\\
\end{aligned}
\end{equation}
eq. \eqref{N1} becomes the following system of coupled partial differential equations:
\begin{equation} \label{N3}
\begin{aligned}
&iU_{t}+U_{xx}+UV=0,\\
&V_{y}=[U(x,y,t)U^{*}(-x,-y,t)]_{x},\\
\end{aligned}
\end{equation}
where $U$ and $V$ are two real functions that satisfy the two-dimensional [(2d)] $\mathcal{PT}$-symmetry condition  $V(x,y,t)=V^{*}(-x,-y,t)$.

The organization of this paper is as follows. In sec. 2, $N$-soliton solutions are derived by employing the Hirota bilinear method. In sec. 3, three kinds of semi-rational solutions are generated by taking the long wave limit of a part of exponential functions in the general expression of the $N$-soliton solution obtained with Hirota method.
The main results of the paper are summarized in sec. 4.

\section{$N$-soliton solution of the  $(2+1)d$ nonlocal Fokas system}
The bilinear forms of eq. \eqref{N1} have been given \cite{we5}
\begin{equation}\label{N4}
\begin{aligned}
&(D^{2}_{x}+i D_{t})g \cdot f =0,\\
&(D_{x}D_{y}+1)f \cdot f =g g^{*},
\end{aligned}
\end{equation}
through the dependent variable transformation:
\begin{equation}\label{N5}
\begin{aligned}
U=g/f,\qquad  V=2(logf)_{xx},\\
\end{aligned}
\end{equation}
where $D$ is the Hirota's bilinear differential operator, and $f$ and $g$ are functions of $x$, $y$, and $t$, subject to the
condition:
\begin{equation}
f(x,y,t)=f^{\ast }(-x,-y,t).  \label{N6}
\end{equation}%

\begin{figure}[!htbp]
\centering
\subfigure[$t=0$]{\includegraphics[height=3cm,width=7cm]{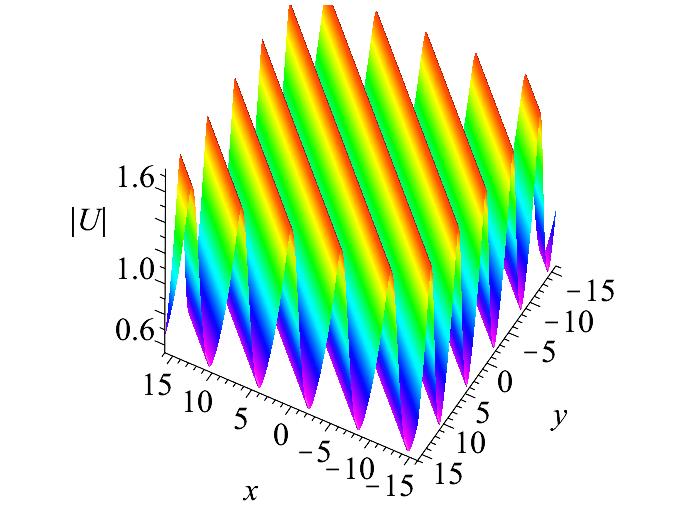}}
\subfigure[$t=0$]{\includegraphics[height=3cm,width=7cm]{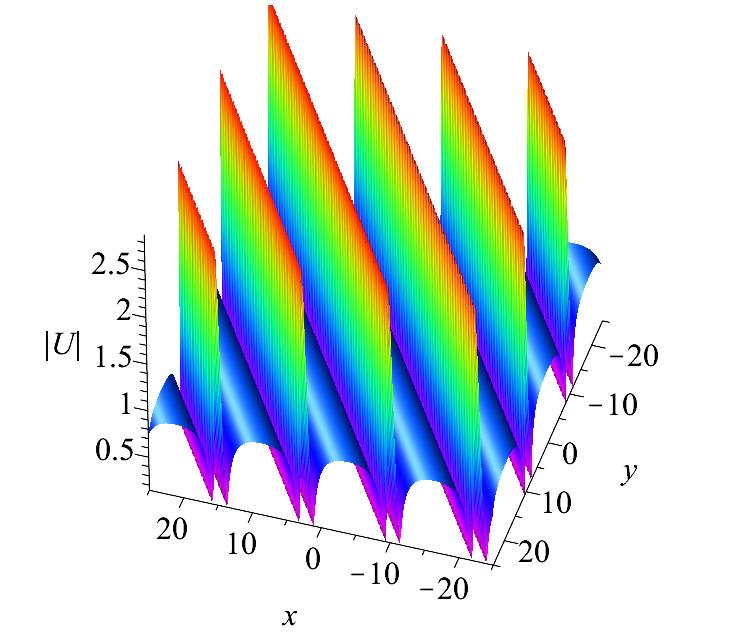}}
\caption{$1$-soliton and $2$-soliton solutions $|U|$ of  eq. \eqref{N1} in the $(x,y)$-plane at $t=0$.}\label{figsoliton}
\end{figure}

Using \eqref{N4} and \eqref{N5}, the $N$-soliton solutions $U$ and $V$ of the $(2+1)d$ nonlocal Fokas system can be obtained by the bilinear transform method \cite{hirota}, in which $f$ and $g$ are written in the following forms:
 \begin{equation}\label{N7}
\begin{aligned}
f=\sum_{\mu=0,1}\exp(\sum_{j<k}^{(N)}\mu_{j}\mu_{k}A_{jk}+\sum_{j=1}^{N}\mu_{j}\eta_{j}),\; \;
g=\sum_{\mu=0,1}\exp(\sum_{j<k}^{(N)}\mu_{j}\mu_{k}A_{jk}+\sum_{j=1}^{N}\mu_{j}(\eta_{j}+i\Phi_{j})).
\end{aligned}
\end{equation}
Here
\begin{equation}\label{N8}
\begin{aligned}
&\Omega_{j}=-\frac{P_j\sqrt{-P_{j}^{2} Q_{j}^{2}+2P_{j}Q_{j}}}{Q_{j}},\eta_{j}=iP_{j}x+iQ_{j}y+\Omega_{j}t+\eta^{0}_{j}, \sin(\Phi_{j})=\sqrt{-P_{j}^{2} Q_{j}^{2}+2P_{j}Q_{j}},\\
&\cos(\Phi_{j})=-P_{j}Q_{j}+1,  \; exp(A_{jk})=\frac{\cos(\Phi_{j}-\Phi_{k})+(Q_{j}-Q_{k})(P_{k}-P_{j})+1}{-\cos(\Phi_{j}+\Phi_{k})+(Q_{j}+Q_{k})(P_{k}+P_{j})-1},
\end{aligned}
\end{equation}
where $P_{j},Q_{j}$ are arbitrary real parameters, $\eta^{0}_{j}$ is a complex constant, and
 the subscript $j$ denotes an integer. The notation $\sum_{\mu=0}$ indicates summation over all possible combinations of $\mu_{1}=0,1,\mu_{2}=0,1,\cdots,\mu_{n}=0,1$.  The $\sum\limits_{j<k}^{(N)}$ summation is over all possible combinations of the $N$ elements with the specific condition $j<k$.  To illustrate the above formula of the $N$-soliton solution, the profiles of the $1$-soliton and $2$-soliton solutions are plotted in Fig. \ref{figsoliton}  according to  eqs. (\ref{N5}) and (\ref{N7}).  They are  periodic line waves, which are the basic bricks for the construction of the semi-rational solutions.

\section{Semi-rational solutions of the  $(2+1)d$ nonlocal Fokas system}
In this section, we focus on the semi-rational solutions of eq. \eqref{N1}. Following our earlier works \cite{rao1,rao2}, the semi-rational solutions can be derived by taking a long wave limit of a part of exponential functions in $f$ and $g$ given by \eqref{N7}.
We obtain three generic types of the so-called hybrid solutions involving (a) breathers, (b) lumps, (c) RWs, and (d) periodic line waves. Thus the hybrid solution is a combination (i.e. a mixture) of two or more of the above mentioned
four distinct types of solutions of the nonlinear equation under consideration.
Indeed, setting the parameters in \eqref{N7}
\begin{equation} \label{N9}
\begin{aligned}
 Q_{k}=\lambda_{k}P_{k}, \eta^{0}_{k}=i\pi, 0<2j<N,1\leq k \leq 2j,\\
\end{aligned}
\end{equation}
and taking the limit $P_{k}\rightarrow 0$, then the functions $f$ and $g$ defined in \eqref{N7} become a combination of polynomial and exponential functions, and they generate semi-rational solutions $U$ and $V$ of eq. \eqref{N1}.
Next, we study three kinds of semi-rational solutions of the $(2+1)d$ nonlocal Fokas system.

\subsection{A hybrid of RWs and periodic line waves}
\begin{figure}[!htbp]
\centering
\subfigure[t=-8]{\includegraphics[height=2.8cm,width=3.2cm]{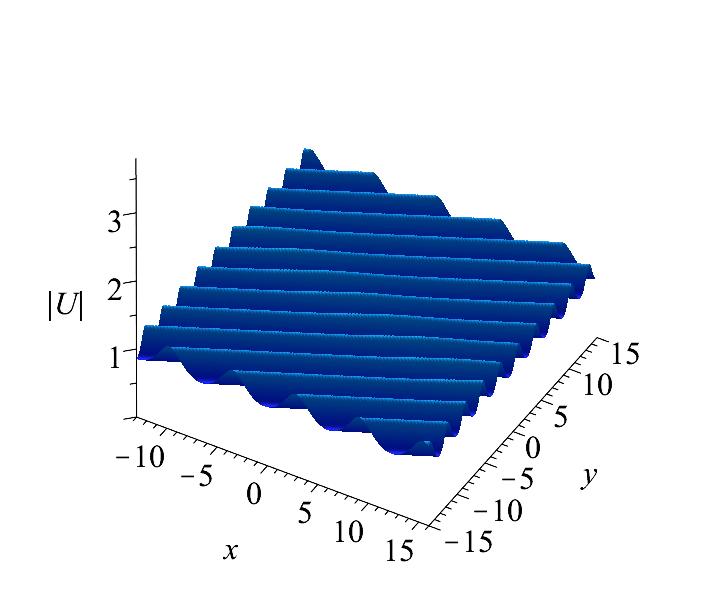}}
\subfigure[t=-2.5]{\includegraphics[height=2.8cm,width=3.2cm]{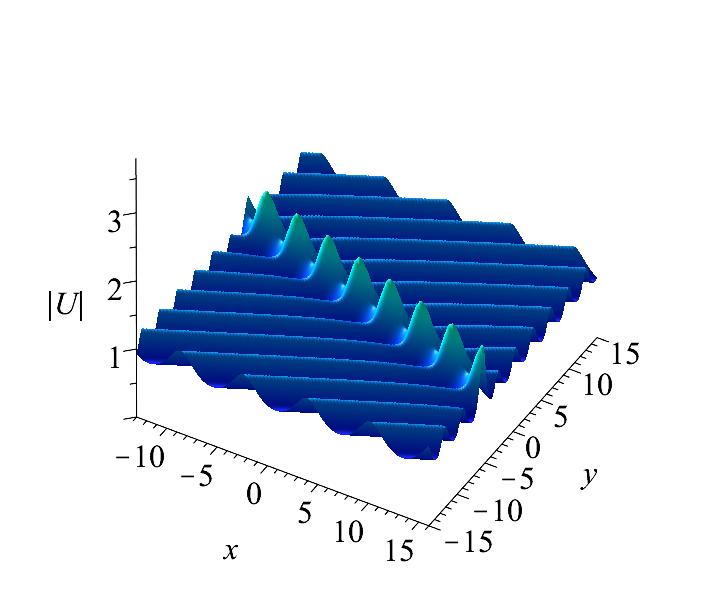}}
\subfigure[t=0]{\includegraphics[height=2.8cm,width=3.2cm]{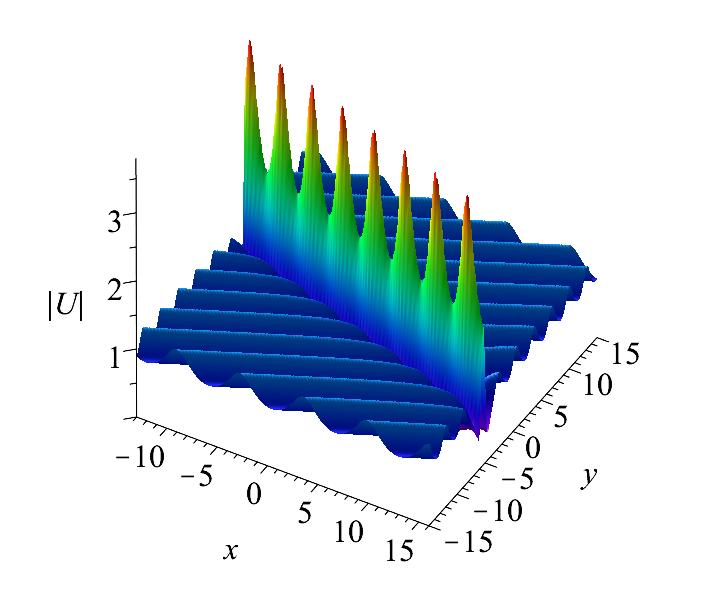}}
\subfigure[t=1.5]{\includegraphics[height=2.8cm,width=3.2cm]{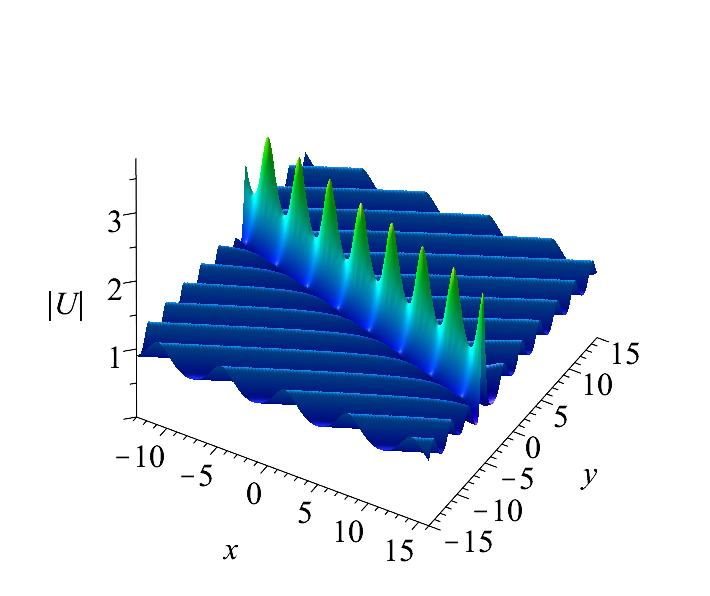}}
\subfigure[t=8]{\includegraphics[height=2.8cm,width=3.2cm]{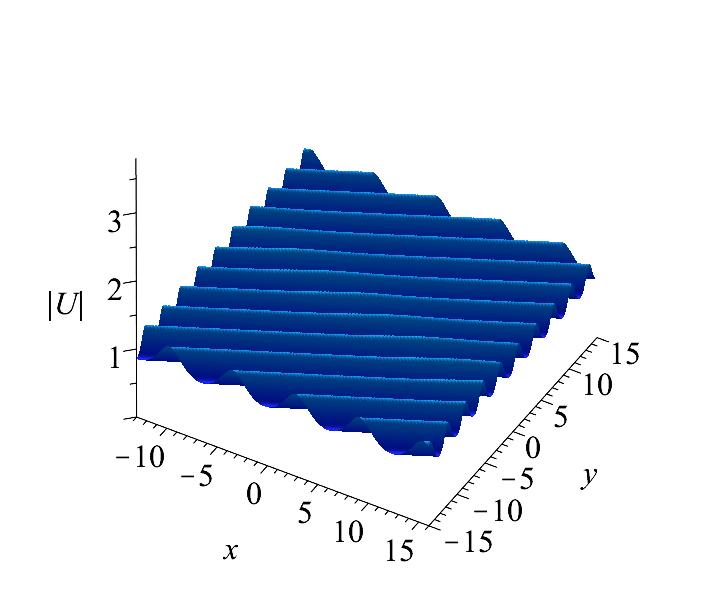}}
\caption{The time evolution in the $(x,y)$-plane of line RW on top of a background formed by periodic line waves with parameters given by eq. \eqref{N17}.}\label{fig1}
\end{figure}

We consider the first type of hybrid solutions consisting of first-order RWs and periodic line waves obtained from the $3$-soliton solutions. For simplicity, setting
\begin{equation}\label{N10}
\begin{aligned}
N=3\,,Q_{1}=\lambda_{1}P_{1}\,,Q_{2}=\lambda_{2}P_{2}\,,\eta_{1}^{0}=\eta_{2}^{0}=i\,\pi\,,
\end{aligned}
\end{equation}
and $P_{1}\,,P_{2}\rightarrow 0$,
then the functions $f$ and $g$ can be rewritten as
\begin{equation} \label{N17}
\begin{aligned}
f=&[-(x+3y+\frac{12i}{5})^2-\frac{2}{3}t^2-\frac{3}{2}]e^{ix+2iy-\frac{\pi}{2}}-(x+3y)^2-\frac{2}{3}t^2-\frac{3}{2},\\
g=&[(x+3y+\frac{12i}{5})^2+\frac{2}{3}(t-3i)^2+\frac{3}{2}]e^{ix+2iy-\frac{\pi}{2}}-(x+3y)^2-\frac{2}{3}(t-3i)^2-\frac{3}{2},\\
\end{aligned}
\end{equation}
where $\lambda_1=\lambda_2=3,P_3=1,Q_3=2,\eta_3^0=-\frac{\pi}{2}$.
The corresponding semi-rational solution $|U|$ is shown in Fig. \ref{fig1}. It is seen that this semi-rational solution describes a line rogue wave on a background of periodic line waves. When $t\rightarrow \pm \infty$, the line rogue wave approaches a  background formed  by the periodic line waves (see the panels at $t=\pm 8$).  In the intermediate time $t=-2.5$, the line rogue wave arises from the periodic  background,  then  the strong interaction around $t=0$  between the line rogue wave and the periodic line waves results in  the occurence of the  higher peaks along
  the line rogue wave, which can reach four times the constant background amplitude.
  Next, along the evolution in time, all peaks on line rogue wave are merged gradually into the
  periodic line waves. Finally, there just exists a periodic  background formed by the periodic line waves when $t\rightarrow \infty$. This provides a different dynamic property of the solution of eq. \eqref{N1} by comparing it with the corresponding solution on a constant background in the case of the nonlocal DS I equation \cite{rao1} (see Fig. (12) therein).

Higher-order semi-rational solutions composed of several RWs and periodic line waves can also be generated in a similar way for larger $N$ in \eqref{N7}. For example, a hybrid solution of two RWs and periodic line waves can be derived from the $5$-soliton solution. Setting
\begin{equation} \label{G1}
\begin{aligned}
N=5\,,Q_{k}=\lambda_{k}P_{k}\,,\eta_{k}^{0}=i\pi\,(\,k=1,2,3,4\,)\,
\end{aligned}
\end{equation}
and letting $P_{i}\rightarrow 0$ in eq. \eqref{N7}, then the functions $f$ and $g$ become
\begin{equation} \label{G2}
\begin{aligned}
f=&(\theta_{1}\theta_{2}\theta_{3}\theta_{4}+a_{12}\theta_{3}\theta_{4}+a_{13}\theta_{2}\theta_{4}+a_{14}\theta_{2}\theta_{3}+a_{23}\theta_{1}\theta_{4}
+a_{24}\theta_{1}\theta_{3}+a_{34}\theta_{1}\theta_{2}+a_{12}a_{34}+\\
&a_{13}a_{24}+a_{14}a_{23})+e^{\eta_{5}}[\theta_{1}\,\theta_{2}\,\theta_{3}\,\theta_{4}+a_{45}\,\theta_{1}\,\theta_{2}\,\theta_{3}+a_{35}\,\theta_{1}\,\theta_{2}\,\theta_{4}
+a_{25}\,\theta_{1}\,\theta_{3}\,\theta_{4}+a_{15}\theta_{2}\,\theta_{3}\,\theta_{4}+\\
&(a_{35}a_{45}+a_{34})\theta_{1}\,\theta_{2}+(a_{25} a_{45}+a_{24})\,\theta_{1}\,\theta_{3}+(a_{25}a_{35}+a_{23})\,\theta_{1}\,\theta_{4}+(a_{15}a_{45}+a_{14})\,\theta_{2}\,\theta_{3}\\
&+(a_{15}a_{35}+a_{13})\,\theta_{2}\,\theta_{4}+(a_{15}a_{25}+a_{12})\theta_{3}\,\theta_{4}+(a_{25}a_{35}\,a_{45}+a_{23}a_{45}+a_{25}a_{34}+a_{24}a_{35})\,\theta_{1}\\&
+(a_{15}a_{35}a_{45}+a_{14}a_{35}+a_{13}a_{45}+a_{15}a_{34})\,\theta_{2}+(a_{15} a_{25}a_{45}+a_{14}a_{25}+a_{15}a_{24}+a_{12}a_{45})\,\theta_{3}\\
&+(a_{15}a_{25}a_{35}+a_{15}a_{23}+a_{13}a_{25}+a_{12}a_{35})\,\theta_{4}+a_{12}a_{34}+a_{13}\,a_{24}+a_{14}a_{23}+a_{12}\,a_{35}\,a_{45}+\\
&a_{13}a_{25}a_{45}+a_{14}a_{25}a_{35}+a_{15}a_{24}a_{35}+a_{15}a_{25}a_{34}\,+a_{15}a_{23}a_{45}+a_{15}a_{25} a_{35} a_{45}]\,,\\
g=&[(\theta_{1}+b_{1})(\theta_{2}+b_{2})(\theta_{3}+b_{3})(\theta_{4}+b_{4})+a_{12}(\theta_{3}+b_{3})(\theta_{4}+b_{4})+a_{13}(\theta_{2}+b_{2})(\theta_{4}+b_{4})+\\
&a_{14}(\theta_{2}+b_{2})(\theta_{3}+b_{3})+a_{23}(\theta_{1}+b_{1})(\theta_{4}+b_{4})
+a_{24}(\theta_{1}+b_{1})(\theta_{3}+b_{3})+a_{34}(\theta_{1}+b_{1})(\theta_{2}\\
&+b_{2})+a_{12}a_{34}+a_{13}a_{24}+a_{14}a_{23}]+e^{\eta_{5}+i\phi_{5}}[(\theta_{1}+b_{1})(\theta_{2}+b_{2})(\theta_{3}+b_{3})(\theta_{4}+b_{4})+\\
&a_{45}\,(\theta_{1}+b_{1})(\theta_{2}+b_{2})(\theta_{3}+b_{3})+a_{35}\,(\theta_{1}+b_{1})(\theta_{2}+b_{2})(\theta_{4}+b_{4})
+a_{25}\,(\theta_{1}+b_{1})(\theta_{3}+\\
&b_{3})(\theta_{4}+b_{4})+a_{15}(\theta_{2}+b_{2})(\theta_{3}+b_{3})(\theta_{4}+b_{4})+
(a_{35}a_{45}+a_{34})(\theta_{1}+b_{1})(\theta_{2}+b_{2})+\\
&(a_{25} a_{45}+a_{24})\,(\theta_{1}+b_{1})(\theta_{3}+b_{3})+(a_{25}a_{35}+a_{23})(\theta_{2}+b_{2})(\theta_{4}+b_{4})+(a_{15}a_{45}+a_{14})\\
&\,(\theta_{2}+b_{2})(\theta_{3}+b_{3})+(a_{15}a_{35}+a_{13})\,(\theta_{2}+b_{2})(\theta_{4}+b_{4})+(a_{15}a_{25}+a_{12})(\theta_{3}+b_{3})(\theta_{4}+b_{4})\\
&+(a_{25}a_{35}\,a_{45}+a_{23}a_{45}+a_{25}a_{34}+a_{24}a_{35})\,(\theta_{1}+b_{1})+(a_{15}a_{35}a_{45}+a_{14}a_{35}+a_{13}a_{45}+a_{15}a_{34})\\
&(\theta_{2}+b_{2})+(a_{15} a_{25}a_{45}+a_{14}a_{25}+a_{15}a_{24}+a_{12}a_{45})\,(\theta_{3}+b_{3})+(a_{15}a_{25}a_{35}+a_{15}a_{23}+a_{13}a_{25}\\
&+a_{12}a_{35})(\theta_{4}+b_{4})+a_{12}(a_{34}+a_{35}\,a_{45})+a_{13}(\,a_{24}+a_{25}a_{45})+a_{14}(a_{23}+a_{25}a_{35})+\\
&a_{15}(a_{24}a_{35}+a_{25}a_{34}\,+a_{23}a_{45}+a_{25} a_{35} a_{45})].
\end{aligned}
\end{equation}
Here
\begin{equation} \label{N21}
\begin{aligned}
&a_{jk} =\frac{ 2\lambda_{j}P_{k}Q_{k}}{-\gamma_{j}\sqrt{2\lambda_{j}P_{k}Q_{k}(2-P_{k}Q_{k})}+P_{k}\lambda_{j}+Q_{k}}(\,j=1,2\,,k=3,4\,),\\
&\theta_{j}=\frac{-i\sqrt{2\lambda_{j}}\gamma_{j}t+i\lambda_{j}^{2}y+ix\lambda_{j}}{\lambda_{j}},b_{j}=i\sqrt{2\lambda_{j}}\gamma_{j},\\
&\alpha_{12}=\frac{2\lambda_{1}\lambda_{2}}{\lambda_{1}+\lambda_{2}-2\sqrt{\lambda_{1}\lambda_{2}}\gamma_{1}\gamma_{2}},\gamma_{1}^{2}=1,\gamma_{2}^{2}=1,\\
\end{aligned}
\end{equation}
$b_{j}\,,\eta_{j}\,(\,1\leq k<j\leq4)$ and $\eta_{5}\,,\phi_{5}$ are given by eqs. \eqref{N21} and \eqref{N8}, respectively,  and $a_{j5} =\frac{ 2\lambda_{j}P_{5}Q_{5}}{-\gamma_{j}\sqrt{2\lambda_{j}P_{5}Q_{5}(2-P_5Q_5)} +P_5\lambda_j+Q_5}$.

\begin{figure}[!htbp]
\centering
\subfigure[t=-10]{\includegraphics[height=3cm,width=4cm]{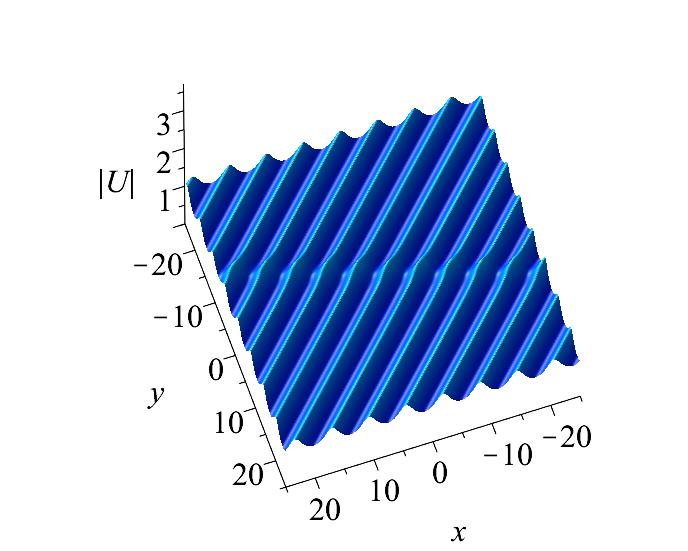}}
\subfigure[t=-2.5]{\includegraphics[height=3cm,width=4cm]{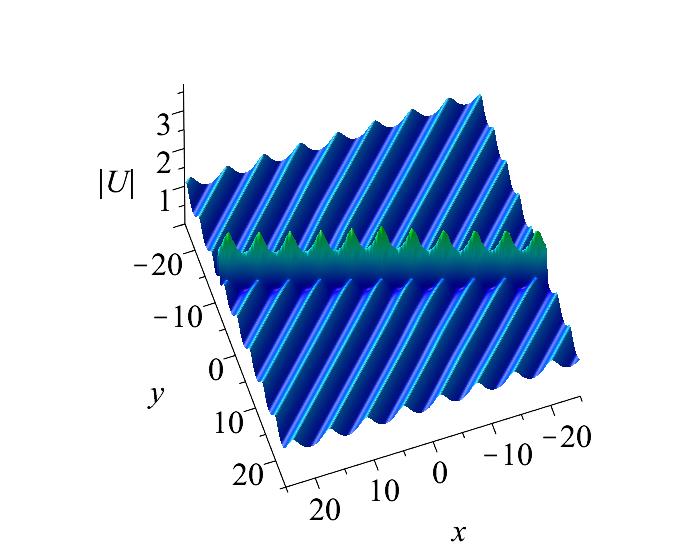}}
\subfigure[t=-0.5]{\includegraphics[height=3cm,width=4cm]{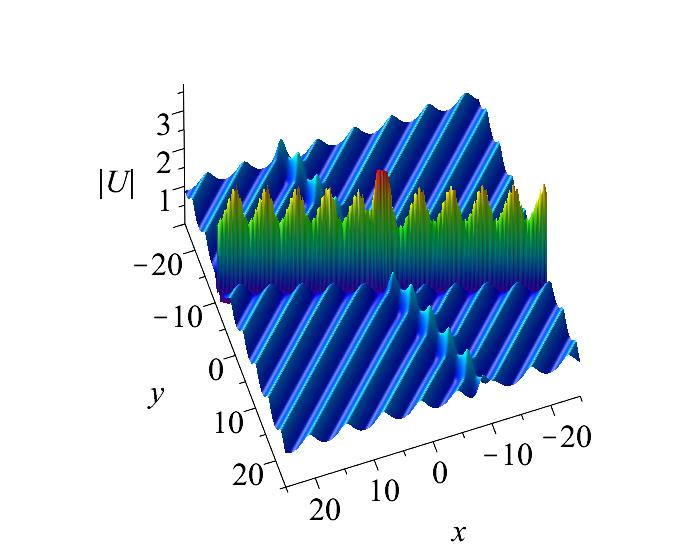}}
\subfigure[t=0]{\includegraphics[height=3cm,width=4cm]{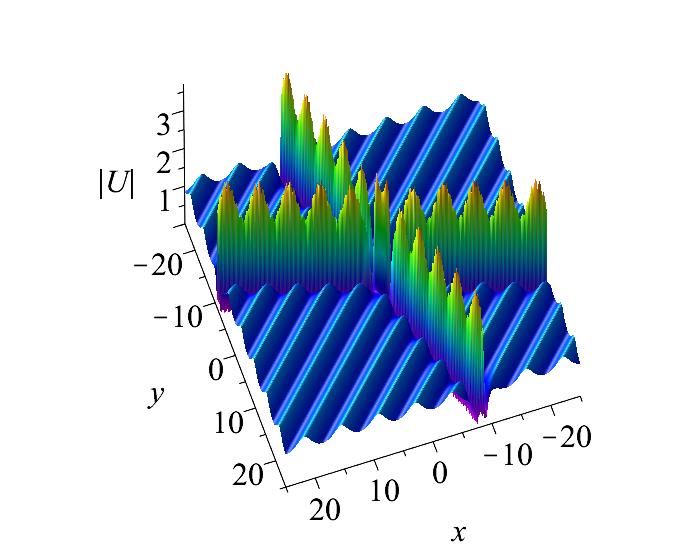}}
\subfigure[t=0.2]{\includegraphics[height=3cm,width=4.5cm]{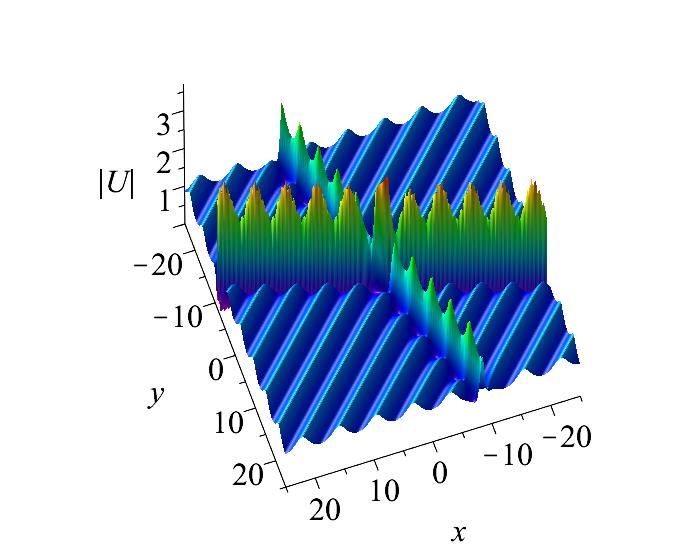}}
\subfigure[t=2]{\includegraphics[height=3cm,width=4.5cm]{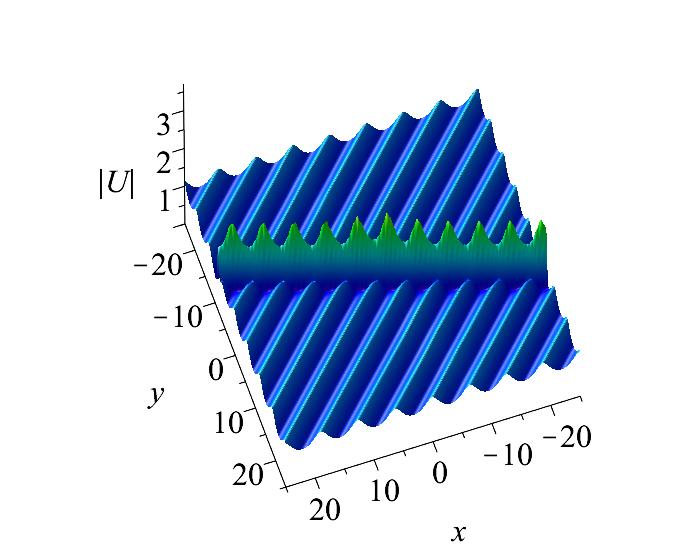}}
\subfigure[t=10]{\includegraphics[height=3cm,width=4.5cm]{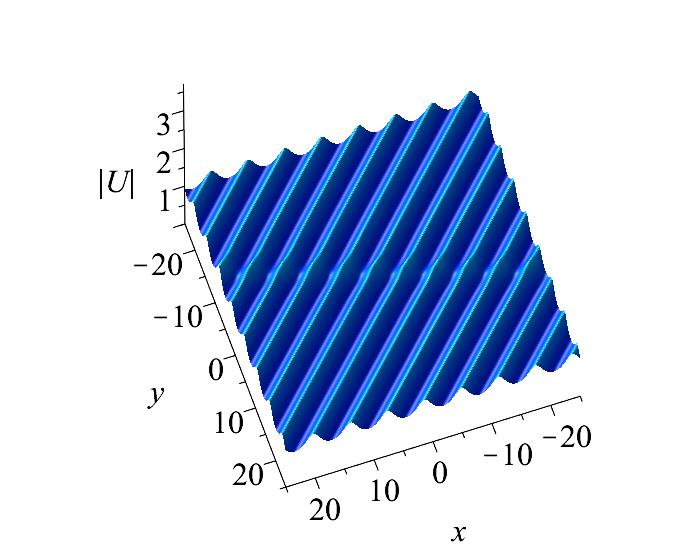}}
\caption{The time evolution in the $(x,y)$-plane of the second-order line RWs on a background of periodic line waves of eq. \eqref{N1} with parameters given by eq. \eqref{G3}.  }\label{fig2}
\end{figure}

Under the parameter constraints
\begin{equation} \label{G3}
\begin{aligned}
\lambda_1=\lambda_2=\frac{1}{3},\lambda_3=\lambda_4=3,P_5=1,Q_5=2,\eta_5^0=-\frac{\pi}{2},
\end{aligned}
\end{equation}
a family of semi-rational solutions $|U|$ are generated, describing the evolution of second-order line RWs on a background of periodic line waves. As can be seen in Fig. \ref{fig2} the seven panels show that two line RWs appear at $t<0$ from a periodic line wave background, gradually attaining a maximum amplitude at intersection area $t=0$, and finally return back to the background of periodic line waves without a trace. This behavior is different from the corresponding dynamics of the semi-rational solutions of the nonlocal DS II equation, see \cite{rao2}.

\subsection{A hybrid of lump and breather}
To construct the second type of semi-rational solution, we have to take the parameters
\begin{equation} \label{N19}
\begin{aligned}
N=4\,,Q_{1}=\lambda_{1}P_{1}\,,Q_{2}=\lambda_{2}P_{2}\,,\eta_{1}^{0}=\eta_{2}^{0}=i\,\pi\,,\eta_{3}^{0}=\eta_{4}^{0},
\end{aligned}
\end{equation}
and take a limit as $P_{1}\,,P_{2}\rightarrow 0$.
Then the functions $f$ and $g$ can be rewritten as
\begin{equation} \label{N20}
\begin{aligned}
f=&e^{A_{34}}(a_{13}a_{23}+a_{13}a_{24}+a_{13}\theta_{2}+a_{14}a_{23}+a_{14}a_{24}+a_{14}\theta_{2}+a_{23}\theta_{1}+a_{24}\theta_{1}+\theta_{1}\theta_{2}\\
&+a_{12})e^{\eta_{3}+\eta_{4}}+(a_{13}a_{23}+a_{13}\theta_{2}+a_{23}\theta_{1}+\theta_{1}\theta_{2}+a_{12})e^{\eta_{3}}+(a_{14}a_{24}+a_{14}\theta_{2}+a_{24}\theta_{1}\\
&+\theta_{1}\theta_{2}+a_{12})e^{\eta_{4}}+\theta_{1}\theta_{2}+a_{12}\,,\\
g=&e^{A_{34}}[a_{13}a_{23}+a_{13}a_{24}+a_{13}(\theta_{2}+b_{2})+a_{14}a_{23}+a_{14}a_{24}+a_{14}(\theta_{2}+b_{2})+a_{23}(\theta_{1}+b_{1})\\
&+a_{24}(\theta_{1}+b_{1})+(\theta_{1}+b_{1})(\theta_{2}+b_{2})+a_{12}]e^{\eta_{3}+i\phi_{3}+\eta_{4}+i\phi_{4}}+[a_{13}a_{23}+a_{13}(\theta_{2}+b_{2})\\
&+a_{23}(\theta_{1}+b_{1})+(\theta_{1}+b_{1})(\theta_{2}+b_{2})+a_{12}]e^{\eta_{3}+i\phi_{3}}+[a_{14}a_{24}+a_{14}(\theta_{2}+b_{2})+a_{24}(\theta_{1}\\
&+b_{1})+(\theta_{1}+b_{1})(\theta_{2}+b_{2})+a_{12}]e^{\eta_{4}+i\phi_{4}}+(\theta_{1}
+b_{1})(\theta_{2}+b_{2})+a_{12},
\end{aligned}
\end{equation}
Here $a_{jk},\theta_{j},\alpha_{12}$ and $\phi_{l},\eta_{l}\,{A_{34}}$ are given by eqs.\eqref{N21} and \eqref{N8},  respectively. Further, we take
\begin{equation} \label{N22}
\begin{aligned}
\lambda_{1}=\lambda_{2}=3,P_{3}=-P_{4}=\frac{1}{2},Q_{3}=-Q_{4}=\frac{1}{4},\eta_{3}^0=\eta_{4}^0.
\end{aligned}
\end{equation}
\begin{figure}[!htbp]
\centering
\subfigure[$\eta_{4}^0=-8\pi$]{\includegraphics[height=3cm,width=3cm]{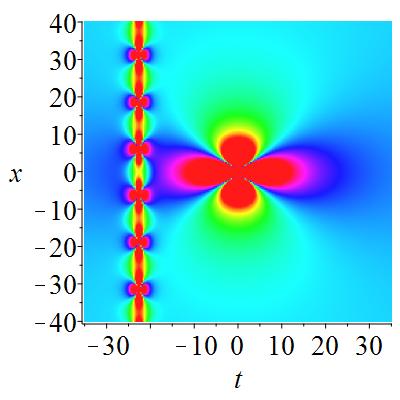}}
\subfigure[$\eta_{4}^0=-2.5\pi$]{\includegraphics[height=3cm,width=3cm]{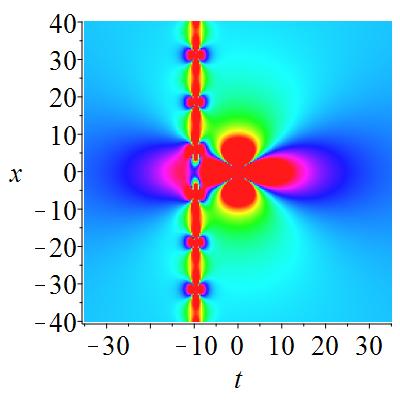}}
\subfigure[$\eta_{4}^0=0$]{\includegraphics[height=3cm,width=3cm]{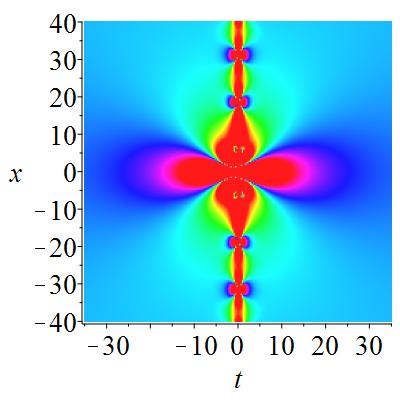}}
\subfigure[$\eta_{4}^0=1.5\pi$]{\includegraphics[height=3cm,width=3cm]{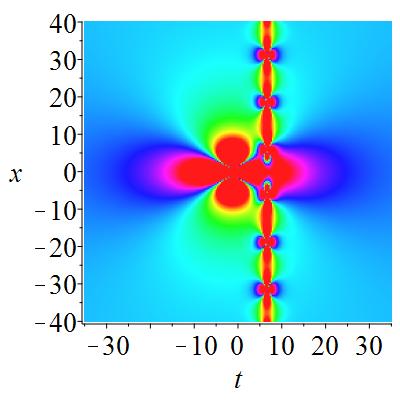}}
\subfigure[$\eta_{4}^0=8\pi$]{\includegraphics[height=3cm,width=3cm]{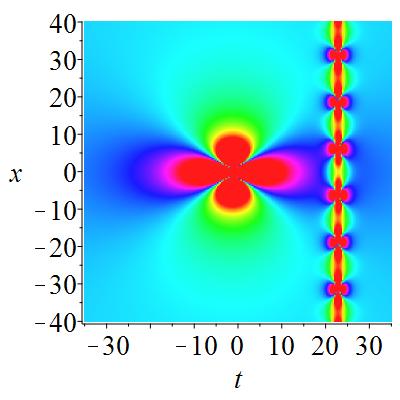}}
\caption{Semi-rational solutions $|U|$ plotted in the $(x,t)$-plane, consisting of lump and Akhmediev breather solutions for eq. \eqref{N1} with parameters given by eq. \eqref{N22}}\label{fig3}
\end{figure}

The semi-rational hybrid solutions $|U|$ as the combination of a first-order lump and of an Akhmediev breather, obtained from eq. \eqref{N20} are plotted in the $(x,t)$-plane, see Fig. \ref{fig3}.  We see that the lump moves and passes through the breather. This unique behavior can be obtained by changing the parameter $\eta_{4}^{0}$. As can be seen, the breather keeps  periodic in the $x$ direction and is localized in space. When $|\eta_{4}^0|>>0$, the lump and breather would separate completely. The amplitude increases at the intersection point, and the period of the breather become smaller, see Fig.  \ref{fig3}(c).

\subsection{A hybrid of lump, breather, and periodic line waves}
\begin{figure}[!htbp]
\centering
\subfigure[y=0]{\includegraphics[height=4cm,width=8cm]{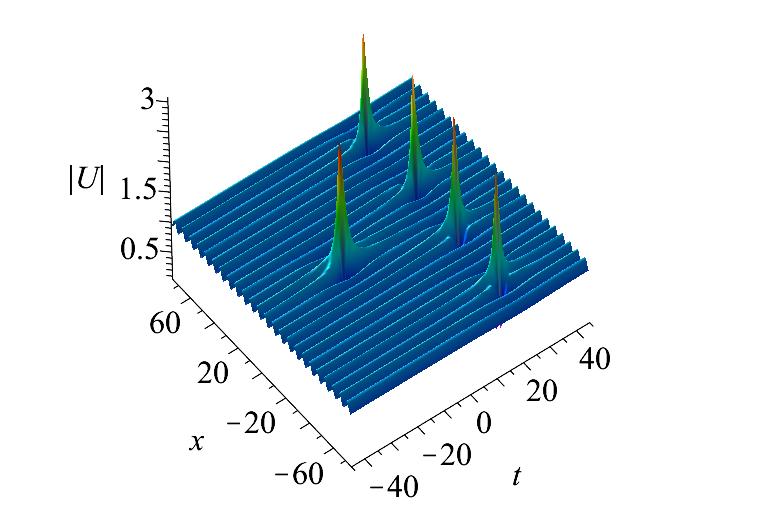}}
\subfigure[y=0]{\includegraphics[height=4cm,width=5cm]{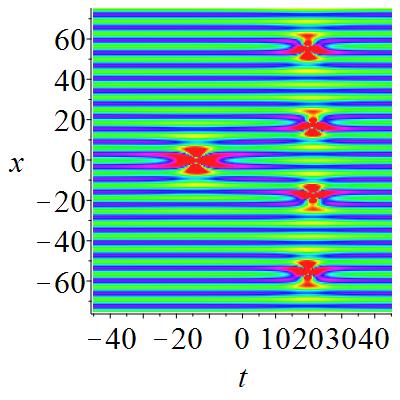}}
\caption{Semi-rational solutions $|U|$ plotted in the $(x,t)$-plane, which consist of lump, Akhmediev breather, and periodic line waves for  eq. \eqref{N1} with parameters $P_3=\frac{1}{6},Q_3=\frac{1}{3},P_5=1,Q_5=2,\eta_{3}^{0}=\eta_{4}^{0}=-\eta_{5}^{0}=\pi,\lambda_1=\lambda_2=3$ in eq. \eqref{Y1}.}\label{fig4}
\end{figure}

We derive here a third type of semi-rational solution composed of a lump, a breather, and periodic line waves for  eq. \eqref{N1}, which could be generated from the $5$-soliton solution. Setting
\begin{equation} \label{Y1}
\begin{aligned}
N=5\,,Q_{1}=\lambda_{1}P_{1}\,,Q_{2}=\lambda_{2}P_{2}\,,\eta_{1}^{0}=\eta_{2}^{0}=i\,\pi\,,P_{3}=-P_{4},Q_{3}=-Q_{4},
\end{aligned}
\end{equation}
and taking a limit as $P_{1}\,,P_{2}\rightarrow 0$ in eq. \eqref{N7}, then the functions $f$ and $g$ are translated into a combination of polynomial and exponential functions. The corresponding semi-rational solution is illustrated in Fig. \ref{fig4};  we see that a fundamental lump and an Akhmediev breather coexist on a background of periodic line waves. Both the breather and the periodic line waves are periodic in space, and the period of the Akhmediev breather is $12\pi$ while that of periodic line waves is $\pi$.

\section{Conclusions}
In this paper, the explicit forms of $N$-soliton solutions and semi-rational solutions of the $(2+1)d$ nonlocal Fokas system, which features the specific $\mathcal{PT}$-symmetry, are derived analytically by employing the Hirota bilinear method. The obtained $1$-soliton and $2$-soliton solutions are series of periodic line waves, see Fig. \ref{figsoliton}. Three kinds of the obtained semi-rational solutions composed of RWs, lumps, breather, and periodic line waves exhibit a range of interesting and rather  complicated dynamics. The hybrid solutions of first-order RW and periodic line waves has been illustrated in detail in  Fig. \ref{fig1}, and the interaction between higher-order RWs and periodic line waves has also been shown, see Fig. \ref{fig2}. Furthermore, by using the $4$-soliton solutions, the dynamics of the superposition between the lump solution and the Akhmediev breather is illustrated in Fig. \ref{fig3}.  Also, a new kind of semi-rational solutions consisting of lump solution, Akhmediev breather, and periodic line waves is proposed, which describes the dynamics of a lump and an Akhmediev breather interacting with  periodic line waves, see Fig. \ref{fig4}.
Our study of semi-rational solutions of the $(2+1)d$ nonlocal Fokas system may be helpful to promote a deeper understanding of nonlinear phenomena in other $\mathcal{PT}$-symmetric systems. Furthermore, the technique of obtaining different types of semi-ratioal solutions reported in this work might be also used in other relevant nonlinear dynamical systems.

\section{Acknowledgments}
This work is supported by the NSF of China under Grant No. 11671219 and the K.C. Wong Magna Fund in Ningbo University.




\section*{References}

\bibliographystyle{elsarticle-num}

\bibliography{vvv_re-modified}






\end{document}